# Kinetics Parameter Optimization via Neural Ordinary Differential Equations

Xingyu Su[a], Weiqi Ji[b], Jian An[c], Zhuyin Ren[a,c,*], Sili Deng[b], Chung K. Law[d,a]

[a] *Center for Combustion Energy, Tsinghua University, Beijing 100084, China*
[b] *Department of Mechanical Engineering, Massachusetts Institute of Technology, Cambridge, MA 02139, USA*
[c] *Institute for Aero Engine, Tsinghua University, Beijing, 100084, China*
[d] *Department of Mechanical and Aerospace Engineering, Princeton University, Princeton, NJ 08544, USA*

**Abstract**

Chemical kinetics mechanisms are essential for understanding, analyzing, and simulating complex combustion phenomena. In this study, a Neural Ordinary Differential Equation (Neural ODE) framework is employed to optimize kinetics parameters of reaction mechanisms. Given experimental or high-cost simulated observations as training data, the proposed algorithm can optimally recover the hidden characteristics in the data. Different datasets of various sizes, types, and noise levels are tested. A classic toy problem of stiff Robertson ODE is first used to demonstrate the learning capability, efficiency, and robustness of the Neural ODE approach. A 41-species, 232-reactions JP-10 skeletal mechanism and a 34-species, 121-reactions n-heptane skeletal mechanism are then optimized with species' temporal profiles and ignition delay times, respectively. Results show that the proposed algorithm can optimize stiff chemical models with sufficient accuracy and efficiency. It is noted that the trained mechanism not only fits the data perfectly but also retains its physical interpretability, which can be further integrated and validated in practical turbulent combustion simulations.

*Corresponding author: Zhuyin Ren,
zhuyinren@tsinghua.edu.cn

# 1. Introduction

Well-developed chemical kinetics models, with satisfying accuracy and conciseness, are essential in the research and design of energy conversion devices and biochemical processes. Classical approaches to building these models include ab initio calculations and reaction templates developed with expert knowledge [1]. Further parameter estimation may use rate rules and other functional group methods. However, these parameter estimates are seldom accurate enough to predict the quantity of interests (QoIs), a refinement process to optimize the kinetics models is often needed. Consequently, in order to meet the accuracy criterion, it is necessary to use experimental data or high-cost data from computational quantum chemistry to optimize the kinetics parameters. On the other hand, with more experimental data accumulated in the past decades, optimizing and updating existing mechanisms with newly acquired data also plays an important role in chemical kinetics research.

The kinetic model optimization process is generally treated as solving an inverse problem, implemented by sensitivity analysis [2,3], uncertainty analysis [4,5], and subsequent data regression [6-8]. Most kinetics optimization approaches first build the response surfaces [4] mapping from the input parameters to the output QoIs, such as ignition delay times (IDTs) and laminar flame speeds. Thus, the output can be obtained at a relatively low cost. Then one can minimize the loss function via optimization algorithms to fine tune the kinetics parameters, i.e., pre-exponential factors, temperature coefficients, and activation energies. The response surface maps the kinetics parameters to a specific QoI, with its form being unlimited. Commonly used mapping functions are polynomial chaos expansion (PCE) [4,6], high dimensional model representation (HDMR) [7], and artificial neural network (ANN) [8]. Once the mapping is constructed with sufficient samples, one can train the low-cost surrogate models with widely used optimization approaches, such as genetic algorithm (GA) [9], stochastic gradient descent (SGD) [10], and Bayesian regression [8,11]. However, building the response surface is often time-consuming while insufficient samples would lead to external non-physical behaviors in the surrogate models [11]. To overcome this problem, efficient modeling of chemical kinetics and affordable gradient calculation hold the potential to directly update the parameters of chemical models.

During the past decade, advances in deep learning have offered opportunities for efficient high-fidelity combustion simulations. The most important advantages include the universal approximation ability, well-developed optimization techniques, and open-source ecosystems. Recently, Rassi et al. [12] proposed the physics-informed neural network (PINN) that allows physical constraint of governing equations during the neural network training process, which enables the distillation of the physical behavior in datasets to control parameters of a given system. Following that, Ji et al. [13] employed PINN in combustion scenarios and proposed the stiff-PINN method to optimize kinetics parameters. There are also plenty of neural network-related studies focusing on building surrogate chemical models [14], constructing sub-grid chemical source terms [15], and discovering unknown reaction pathways [16]. Thus, it is a promising method to utilize neural networks to optimize kinetics mechanisms.

Recently, Chen et al. [17] proposed the neural ordinary differential equation (Neural ODE) approach in deep learning and showed its capability of learning computer vision tasks. It employs MLP layers as the ODE layer and uses neural networks to approximate the given ODE systems or classification tasks by the gradient descent method, with the gradients calculated via the adjoint sensitivity method. The applications of Neural ODE have inspired extensive recent developments in the adjoint sensitivity algorithms [18,19] and associated open-source software ecosystems [20,21]. For example, Ma et al. [19] studied the performance of different implementations of the adjoint sensitivity method and suggested guidelines on the choice of adjoint sensitivity algorithms based on the size and stiffness of the ODE system.

In the present work, we introduce the Neural ODE concept to optimize realistic chemical ODE systems to take advantage of these recent developments. The proposed kinetics parameter optimization framework is integrated into a trackable data structure in the Julia language. The gradient of loss function against model parameters is obtained by the adjoint sensitivity method, which is accelerated by utilizing the Jacobian function from automatic differentiation. Specifically, the ODEs are written as neural networks and then integrated by stiff ODE solvers, and the kinetics parameters are then constrained by the loss function. Meanwhile, the neural network is still interpretable which facilitates providing physical insights and integration of the kinetic model into large-scale turbulent combustion simulations.

The subsequent contents are arranged as follows. In Section 2, the algorithm is detailed and the training progress is briefly introduced, with a toy problem Robertson ODE tested against different noise levels. In Section 3, a JP-10 skeletal mechanism is used to demonstrate the optimization ability of the proposed method. Furthermore, an n-heptane skeletal mechanism is generated and optimized against its parent mechanism under the constraint of ignition delay times. Conclusions are presented in Section 4.

# 2. Methodology

## 2.1 The neural network architecture

Consider a typical dynamics system described by an ordinary differential equation (ODE):

$$\frac{du}{dt} = f(u, t, \theta) \quad (1)$$

where $u$ is the state variable, $t$ the time, and $\theta$ the model parameters that control the dynamics. Given the initial state, this system can be integrated by ODE solver

$$u(t_1) = u(t_0) + \int_{t_0}^{t_1} f(u, t, \theta) dt \quad (2)$$

For simple non-stiff ODE systems such as the Lotka-Volterra model for predator-prey population dynamics [19], one can use explicit ODE solvers, e.g., Runge-Kuta methods. But chemical kinetics of hydrocarbon fuels, in general, involves a wide range of chemical timescales, and the resulting ODEs are highly stiff. The largest and smallest eigenvalues $\lambda_{max}$ and $\lambda_{min}$ of the system's Jacobian matrix can differ in many orders of magnitude. In this study, the implicit second-order backward difference formula with trapezoidal rule (TRBDF2) is employed for integrating stiff chemistry assisted by the interface of the Arrhenius.jl package [21].

In the following, a simple reaction system involving three species [A, B, C] is used to illustrate the network structure in neural ODEs

$$A \xrightarrow{k_1} B; \ B + B \xrightarrow{k_2} C + B; \ B + C \xrightarrow{k_3} A + C \quad (3)$$

where $k_1, k_2, k_3$ are the rate constants of each reaction. Without loss of generality, all the reactions can be written in the form of

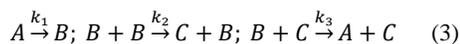
$$v_{i,A}^f A + v_{i,B}^f B + v_{C,i}^f C \xrightarrow{k_i} v_{i,A}^r A + v_{i,B}^r B + v_{i,C}^r C \quad (4)$$

The constants $v_{i,j}^f$, $v_{i,j}^r$ are forward and backward stoichiometric coefficients of species $j$ in the $i$-th reaction. The reaction rate of the $i$-th elementary reaction is described by the power-law expression as

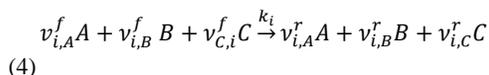
$$r_i = k_i [A]^{v_{i,A}^f} [B]^{v_{i,B}^f} [C]^{v_{i,C}^f} \quad (5)$$

Viewing from the perspective of a neural network, this equation can be formed by a series of linear combinations and activation functions, i.e.,

$$r_i = \exp\left(\ln k_i + v_{i,A}^f \ln[A] + v_{i,B}^f \ln[B] + v_{i,C}^f \ln[C]\right) \quad (6)$$

in which the dependency of $k_i$ on temperature is usually described by the three-parameters Arrhenius formula,

$$k_i = A_i T^{b_i} \exp\left(-\frac{Ea_i}{RT}\right). \quad (7)$$

That is, $\ln k_i = \ln A_i + b_i \ln T - \frac{Ea_i}{RT}$. Note that the reaction rate of species (say [A]) is a linear combination of the reaction rates of elementary reactions, i.e.,

$$\frac{d[A]}{dt} = \sum_i -v_{i,A}^f r_i + v_{i,A}^r r_i. \quad (8)$$

Therefore, the chemical reaction network can be formulated as a neural network that consists of multiple layers of activation functions and linear connections. As shown in Fig. 1, the ODEs of the simple reaction system are transformed to an ODE Layer in the Neural ODE's network. For the neural ODE framework, following ResNet [22], the integration process is achieved by adding the ODE Layer's input to its output, i.e., $u_{i+1} = u_i + \text{ODELayer}(u_i)$. Note that this neural ODE framework is different from the one proposed by Chen et al. [17,23,24]. In [17], the network also consists of ODE layers and is integrated by ODE solvers, but each ODE layer is still made of MLP layers constructed by fully connected neural networks. In this study, the ODE layer is however made of exactly the original ODE equations that are viewed, modeled, and optimized in the neural network perspective.

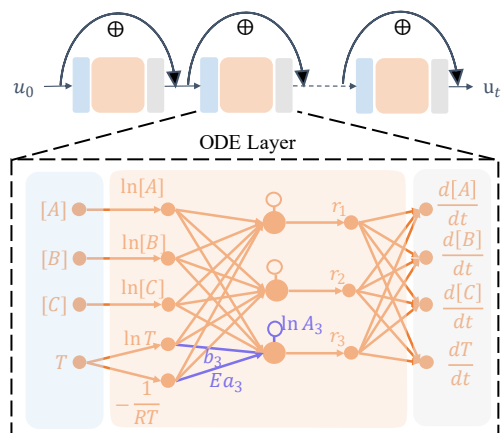

Fig. 1 Schematics of network architecture for neural ordinary differential equations.

Thus, the chemical reaction models are now transformed into a neural network and one can train the learnable parameters with datasets, as well as get the ODE system constrained on given observations. Under this implementation, one can optimize the kinetics parameters with SGD optimizers like Adam [25], once gradients against loss functions are obtained.

*2.2 Adjoint sensitivity method*

Integration of the chemical neural network is usually time consuming when the numbers of species and reactions are large. The integration process often also leads to thousands of ODE Layers mapping, making the backpropagation difficult for the neural ODE. To address this challenge, Chen et al. [17] employed the adjoint method to reduce the memory complexity of gradient computation.

For illustration, let us consider a loss function only concerning the state variables at a single time instant $t_1$:

$$\min L(u,\theta) = L(u(t_1),\theta)$$
$$s.t. \frac{du}{dt} = f(u,t,\theta) \quad (9)$$

To optimize this system, the gradient of loss function against the parameters $\frac{dL}{d\theta}$ is needed. Denoting the adjoint sensitivity of $u$ as $a(t) \equiv -\frac{dL}{du(t)}$. Perturbing time $t$ with sufficient small $\varepsilon$, one has

$$u(t+\varepsilon) = u(t) + \int_t^{t+\varepsilon} f(u,t,\theta)\,dt$$
$$= u(t) + \varepsilon f(u,t,\theta) + \mathcal{O}(\varepsilon^2) \quad (10)$$

$$a(t) = \frac{dL}{du(t)} = \frac{dL}{du(t+\varepsilon)} \frac{du(t+\varepsilon)}{du(t)}$$
$$= a(t+\varepsilon)\left(I + \varepsilon \frac{\partial f(u,t,\theta)}{\partial u(t)} + \mathcal{O}(\varepsilon^2)\right) \quad (11)$$

Therefore, the governing equation for the adjoint sensitivity is

$$\frac{da(t)}{dt} = \lim_{\varepsilon \to 0^+} \frac{a(t+\varepsilon) - a(t)}{\varepsilon}$$
$$= \lim_{\varepsilon \to 0^+} \frac{a(t+\varepsilon) - a(t+\varepsilon)\left(I + \varepsilon \frac{\partial f(u,t,\theta)}{\partial u(t)} + \mathcal{O}(\varepsilon^2)\right)}{\varepsilon} \quad (12)$$
$$= -a(t)\frac{\partial f(u,t,\theta)}{\partial u(t)}.$$

As implemented in neural networks, the Jacobian matrix $\frac{\partial f(u,t,\theta)}{\partial u(t)}$ can be obtained by automatic differentiation of chain rules. Very similarly, denoting the parameter adjoint sensitivity $a_\theta(t) \equiv \frac{dL}{d\theta(t)}$, one can finally obtain the gradient of loss function via backward integration of

$$\frac{dL}{d\theta} = a_\theta(t_0) + \int_{t_1}^{t_0} -a(t)\frac{\partial f(u,t,\theta)}{\partial \theta}\,dt$$
$$a(t_1) = \frac{\partial L}{\partial u(t_1)} + \int_{t_1}^{t_0} -a(t)\frac{\partial f(u,t,\theta)}{\partial u(t)}\,dt \quad (13)$$

The governing ODEs of the adjoint states are usually called augmented dynamics. By only integrating the augmented dynamics, one can obtain the full gradients of the given loss function with respect to the training datasets. As shown in the top panel of Fig. 2, the forward integration will get the state variables of the ODE, and there are discrepancies between state variables and the data represented by symbols. Backward integration of the augmented dynamics results in the adjoint state shown in the bottom panel of Fig. 2. When multiple times are used to constrain a given system, the backward integration takes the errors at each time into account, as demonstrated by the step change of adjoint state at corresponding times.

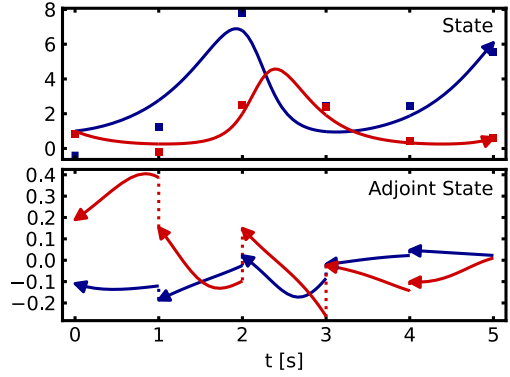

Fig. 2 Demonstration of forward and backward integration of the Lotka-Volterra ODE and its augmented dynamics. Readers can refer to [17,19] for more details.

### 2.3 Kinetics parameter optimization

By integrating the adjoint sensitivity, one can easily get the loss gradients and optimize the kinetics parameters of any chemical model under various types of datasets. As for biomedical processes or chemical pyrolysis processes, time related species profiles are generally measured or estimated in experiments and models are trained to fit the experimental observations. In combustion scenarios, shock tube oxidation data plays a similar role, with much more available data are indirect characteristics of given models, such as ignition delay times (IDTs) or laminar flame speeds.

To demonstrate the proposed framework's capability and robustness to optimize kinetics parameters under different scenarios, a classical model problem, the stiff Robertson ODE [26] is first tested. The Robertson ODE shares the same dynamics with Eq (3) but do not involve temperature-dependent rate constants. The general used rate constants $k^{true} = [0.04, 3 \times 10^7, 1 \times 10^4]$ are adopted to generate the ground truth data with initial value $y_0 = [1,0,0]$. The data samples are drawn from 50 uniform time instants in log time scale of range $[10^{-4}, 10^8]$ s.

Randomly perturbing the kinetics parameter to $k^{init} = [0.023, 1.88 \times 10^8, 1.96 \times 10^3]$, and adding 10% noise to the datasets, one can then train the Robertson ODE with loss function constraining on the state variable $y$. Based on the fact that chemical species concentration span across several orders of magnitude, the loss function utilizes a normalization via $y_{scale}$, which is a vector in the same shape of $y_0$ and each value describes the scale of each state variable, $L(y,\theta) = MSE\left(\frac{y^{model}-y^{obs}}{y_{scale}}\right)$ where $y^{model}$ is the prediction of neural networks, $y^{obs}$ is the observed noisy data, $y_{scale}$ is obtained by subtracting the min from the max of each state variable, and MSE represents the mean squared error function.

As shown in Fig. 3, the predictions with parameters $k_{init}$ differ significantly with the ground true data. After 200 epochs training, the optimized ODE well predicts the data behavior and the optimized parameters $k^{opt} = [0.041, 3.00 \times 10^7, 1.02 \times 10^4]$ is very close to the true ones.

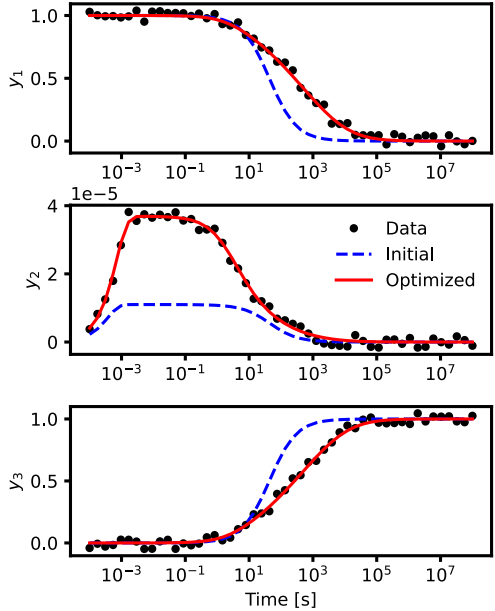

Fig. 3 Training results of Robertson ODE with 10% noise level.

To examine the robustness of the Neural ODE method, another 100 training experiments are performed under noise levels between 0.1% to 20%. Each experiment uses random initial parameters and a fixed 300 epochs training is employed. To quantify the training results, normalized errors of $y$ and $k$ are defined as $y_{err} = \sqrt{\frac{1}{N \cdot D} \sum \left(\frac{y^{opt}-y^{obs}}{y_{scale}}\right)^2}$ and $k_{err} = \sqrt{\frac{1}{D} \sum \left(\frac{k^{opt}-k^{true}}{k_{true}}\right)^2}$, where $N = 50$ is the number of data samples and $D = 3$ is the dimension of state variables. The results of $y_{err}$ and $k_{err}$ are shown in Figure 4. Intuitively, converged training leads to $y_{err}$ being comparable to noise level so $y_{err}$ collapses around the line $y = x$. The $k_{err}$ results show that the learned kinetics parameters indeed remain similar or have smaller scale of error compared to the noise level, which suggest good learning ability and robustness of the Neural ODE method.

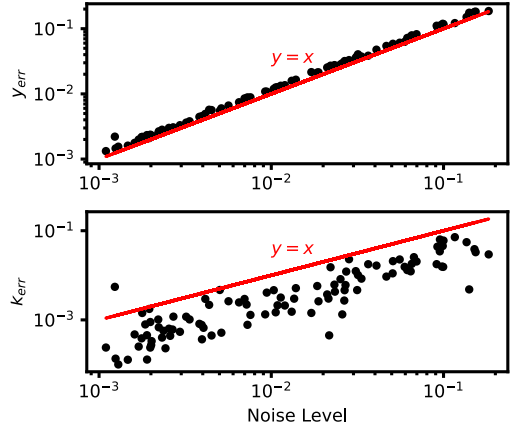

Fig. 4 Errors of Robertson ODE's training results against different noise levels.

## 3. Results and Discussion

In this section, we demonstrate the capability of Neural ODE approach in optimizing the kinetics parameters in a JP-10 skeletal mechanism with known species pyrolysis profiles and for an over-reduced *n*-heptane mechanism with known ignition delay times from detailed one.

### 3.1 Case 1: JP-10 pyrolysis

Jet propellant 10, or JP-10, is one of the leading high volumetric energy density fuel candidates for propulsion devices. Tao et al. [27] developed a 41-species, 232-reaction mechanism for JP-10 surrogate $C_{10}H_{16}$ by constraining this hybrid chemistry model with high temperature pyrolysis experimental data. Here by artificially perturbing the kinetics parameters e.g., rate constants $k$, the test aims to demonstrate if Neural ODE can recover the original ones, with the dataset of species pyrolysis profiles being generated with the Tao mechanism. The datasets consist of 20 training sets with initial temperature range of 1000-1200K, initial mass fraction of $C_{10}H_{16}$ in the range of 0.02-0.2; and another 5 validation sets with initial temperature in 1200-1400K and initial $Y_{C_{10}H_{16}}$ in 0.2-0.3. The true parameters $k_{true}$ are the original Arrhenius parameters proposed by Tao et al. [27] By defining $\theta = \exp(k/k_{true})$, the true kinetics parameters are hence represented by $\theta_{true} = \mathbf{0}$. The datasets are obtained by solving the chemical ODEs with $k_{true}$ and sampling 10 points uniformly in the range of $10^{-6}$ s – $10^{-1}$ s in log scale.

The loss function for training has a MSE term and a regularization term as

$$L(y,\theta) = MSE\left(\frac{y^{model}-y^{obs}}{y_{scale}}\right) + \alpha\theta^T\theta \quad (14)$$

where $\alpha$ is a small number, e.g., $1 \times 10^{-4}$ to avoid unnecessary parameter changes. During the training process, the 20 training sets are randomly permutated and employed to compute gradients in each epoch. To accelerate the training process, larger ODE tolerances and fewer timesteps are used at the beginning of training and gradually adjusted to finer tolerances and timesteps.

Given random initial parameters in the range of $[-1, 1]$, after 100 epochs training, the parameters are well optimized to capture the evolution characteristics of data. As shown in Fig. 5, the initial parameters $\theta_{init}$ result in significant errors in the pyrolysis process. After training, all the species correlate perfectly with the ground truth data. Note that the validation sets are not exposed to the training program, which means the algorithm learned the inner kinetics from training sets and it applies to other thermochemical conditions.

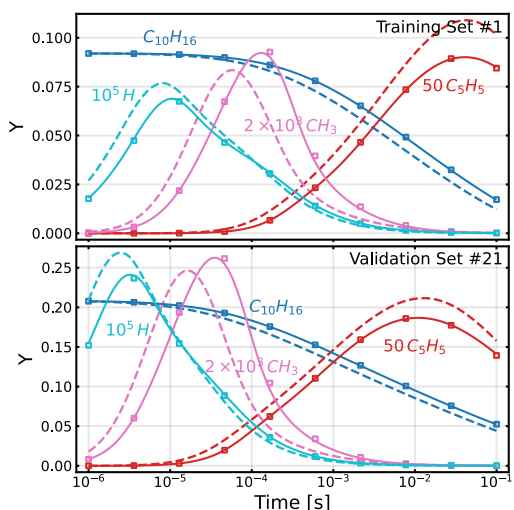

Fig. 5 Training results of JP-10 pyrolysis with temperature and all species profiles as training data. Squares: Data, dash lines: Initial, solid lines: Optimized.

One can further check if the parameters are trained to the true kinetics parameters used in datasets. As shown in Fig. 6, the ground truth $\theta_{true} = \mathbf{0}$ for every elementary reaction, and the initial parameters $\theta_{init}$ randomly distribute in the range $[-1, 1]$, which means around 2.7 times smaller or larger than the original rate constants $k_{true}$. In contrast, for the sensitive reactions, most optimized parameters $\theta_{opt}$ collapse to the x-axis, which means they are optimized toward the ground truth ones. This further confirms that the optimization algorithm extracts the inner kinetics behavior from datasets and gets the parameters optimized to the real ones effectively. Figure 7 further demonstrates the capability of the optimization algorithm by showing that it can well reproduce the pyrolysis process of all species even if only the profiles of a small subset of species are available for training. Note that the information of species $C_5H_5$, $CH_3$ and $H$ are not available, but their evolutions are well reproduced after the kinetics parameters are optimized. The demonstration reveals that it is possible to learn the kinetic parameters from partially observed species profiles. For instance, one can employ the approach on synthesized data to identify the most influential measurements on the kinetic parameter optimization and guide the design of experiments.

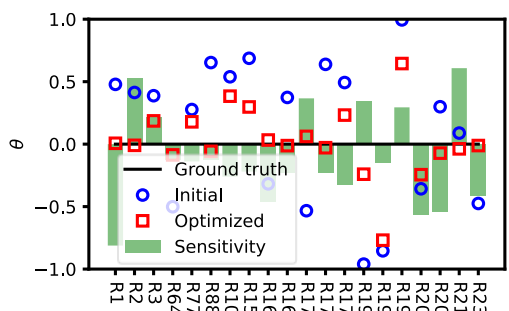

Fig. 6 The initial and optimized kinetics parameters with the ground truth being all zero. Only the most sensitive 21 reactions are plotted.

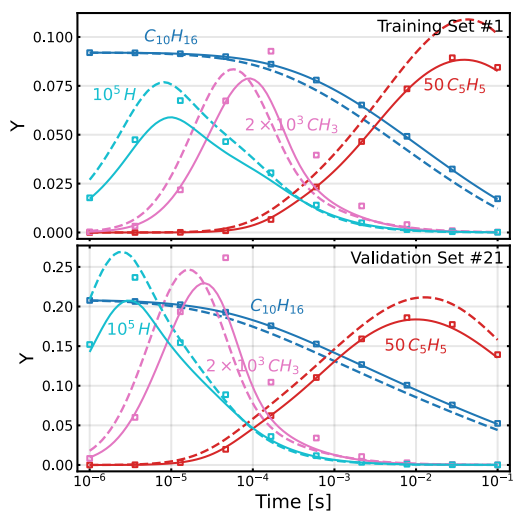

Fig. 7 Training results of JP-10 pyrolysis with only the profiles of $CH_4$, $C_2H_4$ as training data. Squares: Data, dash lines: Initial, solid lines: Optimized.

### 3.2 Case 2: n-heptane autoignition

Most skeletal chemical models are obtained by removing unimportant species and the associated reactions from the master model, leaving the kinetic parameters of the remaining pathways unchanged after the reduction. There has been increasing interest in optimizing the kinetic parameters in overly reduced reaction models to compensate for the error introduced by over-reduction [9,28]. Thus, one can obtain a smaller model with higher fidelity than those acquired via traditional reduction methods.

Here a widely used 41-species, 168-reaction n-heptane mechanism developed by Nordin et al. [29] is employed to demonstrate the ability of neural ODE for optimizing reduced models. That is species $C_3H_5$, $C_3H_4$, $C_2H_6$, $CH_4O_2$, $CH_3O_2$, and $C_2H_2$ are removed via an iterative reduction process. The overly reduced model is denoted as SK34 with 34 species and 121 reactions. It is noted that one can also optimize an existing empirical semi-global reaction model against a detailed model without consulting skeletal reduction.

Ignition delay times (IDTs) are used as the single target characteristics to optimize the skeletal mechanism SK34, with all the parameters including $A$, $b$, $Ea$ being mutable. The training sets $\tau^{obs}$ are IDTs under 500 thermochemical conditions sampled in the range of 700-1600K for initial temperature, 1-50atm for pressure, and 0.5-2.0 for the equivalence ratio. Splitting the datasets with 80% as the training set and 20% as the validation set, one can optimize the problem with the training set and check the validity and reliability with the validation set. The loss function adopted is also an MSE term accompanied with a regularization term:

$$L(\theta) = \left(\log \frac{\tau}{\tau^{obs}}\right)^2 + \alpha \theta^T \theta \quad (15)$$

where $\alpha = 1 \times 10^{-4}$. The gradient of $L(\theta)$ is

$$\frac{dL}{d\theta} = \frac{\partial L}{\partial \tau}\frac{d\tau}{d\theta} + \frac{\partial L}{\partial \theta} = 2\left(\log\frac{\tau}{\tau^{obs}}\right)\frac{d\tau}{dT}\frac{dT}{d\theta} + 2\alpha\theta \quad (16)$$

where $\frac{dT}{d\theta}$ are obtained via the adjoint sensitivity method and $\frac{d\tau}{dT}$ are obtained after the integration of the ODE solver (by numerical gradient evaluation from temperature profile).

In each epoch, 40 samples are randomly chosen from the training set and the gradients are fed to the Adam [25] optimizer with an initial learning rate of $2 \times 10^{-3}$ and default weight decay parameters of 0.9 for the 1st-order moment and 0.999 for the 2nd-order moment. After 100 epochs training, as shown in Fig. 8, the skeletal mechanism is optimized and its prediction on IDTs matches well with the original Nordin mechanism. As for the computational efficiency, the entire optimization process with 363 mutable parameters takes about 10 CPU hours on a normal PC. For reference, the optimization of an overly reduced Jet A kinetic model with 30 mutable parameters among 254 reactions using evolutionary algorithms has to rely on high-performance clusters and cost around hundreds of CPU hours [9]. Figure 9 further validates the evolution of species and temperature during the autoignition process. The optimized skeletal mechanism can not only well predict IDTs and temperature evolution, but also preserve the physical interpretability of the original mechanism, which leads to correct predictions of minor species evolution. Similar observations exist for other species (not shown).

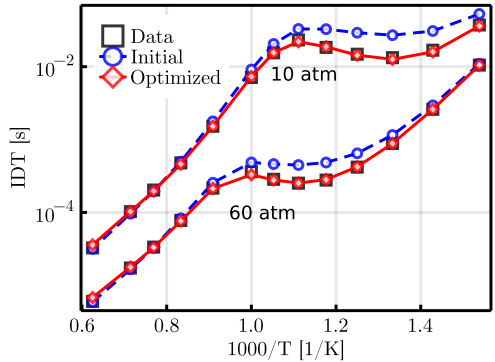

Fig. 8 Prediction on IDTs of Nordin (Data) with its skeletal SK34 (Initial) and the optimized mechanisms SK34OP (Optimized).

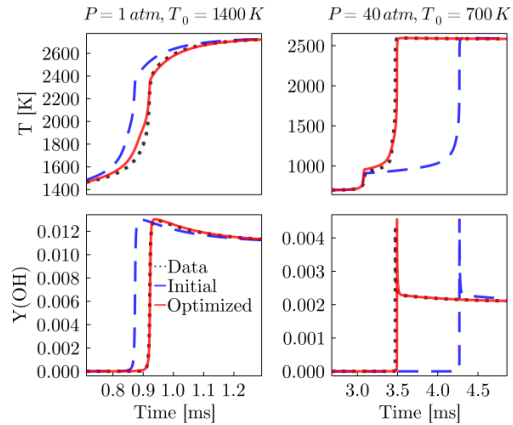

Fig. 9 Profiles of temperature and species OH during autoignition, where the left panels show high-temperature pathway and the right panels for the low-temperature pathway.

## 4. Conclusion

In this work, the Neural ODE architecture with the adjoint sensitivity method is proposed for kinetics parameter optimization with efficient and accurate gradient evaluation of chemical ODE models.

The numerical experiments of optimizing the Robertson problem show that the proposed method can optimize kinetics parameters effectively and robustly, even with a substantial level of noise. The case study of JP-10 pyrolysis demonstrates the ability of the proposed method to learn the true kinetics parameters instead of being trapped into a local minimum, in practice chemical models. Though few data are exposed to the optimization process, the kinetics parameters are adequately optimized and their predictions well fit the training sets and validation sets. Additionally, the parameter optimization of the *n*-heptane skeletal mechanism shows that reduced models can be trained to perform as well as the detailed ones, while not only matching the datasets constraint but also preserving the physical interpretability.

The proposed method can also be employed to optimize global reaction mechanisms, with available experimental data in species concentrations or laminar flame speeds. Assisted with high-efficiency neural network platforms and specific acceleration hardware units, e.g., TPU, one can optimize even larger kinetics models with thousands of reactions.

## Acknowledgments

The work was supported by the National Natural Science Foundation of China 52025062 and National Science and Technology Major Project 2017-III-0004-0028.